\documentclass[11pt,a4paper]{article}

\usepackage[noblocks]{authblk}
\usepackage[utf8]{inputenc}
\usepackage[margin=1in]{geometry}
\usepackage{epsfig}
\usepackage{hyperref} 
\usepackage{physics}
\usepackage{breqn}
\usepackage{multirow}
\usepackage{lscape}
\usepackage{tablefootnote}

\title{Electronic structure of Rf$^+$ (Z = 104) from \textit{ab initio} calculations}

\author[1,2]{Harry Ramanantoanina\footnote{Corresponding author: haramana@uni-mainz.de}}
\author[3]{Anastasia Borschevsky}
\author[1,2,4]{Michael Block}
\author[1,2]{Mustapha Laatiaoui}

\affil[1]{Department Chemie, Johannes Gutenberg-Universit\"at, Fritz-Strassmann Weg 2, 55128 Mainz, Germany}
\affil[2]{Helmholtz-Institut Mainz, Staudingerweg 18, 55128 Mainz, Germany}
\affil[3]{Van Swinderen Institute for Particle Physics and Gravity, University of Groningen, Nijenborgh 4, 9747 Groningen, The Netherlands}
\affil[4]{GSI Helmholtzzentrum f\"ur Schwerionenforschung, Planckstrasse 1, 64291 Darmstadt, Germany}

\date{\today}

\begin{document}
\maketitle

\begin{abstract}
We report calculation of the energy spectrum and the spectroscopic properties of the superheavy element ion: Rf$^+$. We use the 4-component relativistic Dirac-Coulomb Hamiltonian and the multireference configuration interaction (MRCI) model to tackle the complex electronic structure problem that combines strong relativistic effects and electron correlation. We determine the energies of the ground and the low-lying excited states of Rf$^+$, which originate from the 7\textit{s}$^2$6\textit{d}$^1$, 7\textit{s}$^1$6\textit{d}$^2$, 7\textit{s}$^2$7\textit{p}$^1$, and 7\textit{s}$^1$6\textit{d}$^1$7\textit{p}$^1$ configurations. The results are discussed vis-à-vis the lighter homologue, Hf$^+$ ion. We also assess the uncertainties of the predicted energy levels. The main purpose of the presented calculations is to provide a reliable prediction of the energy levels and to identify suitable metastable excited states that are good candidates for the planned ion-mobility-assisted laser spectroscopy studies.\par
\end{abstract}

\section{Introduction}
With the recent confirmation of four new elements \cite{eichler2019,schwerdtfeger2020}, the seventh row of the periodic table is now complete. Superheavy elements (SHE), with atomic number Z $>$ 103, are part of the seventh period. They do not occur on earth but are synthesized in single atom-at-a-time quantities \cite{laatiaoui2019a,block2021}. Moreover, they are short-lived, so that their experimental investigation is also very challenging. Recently, within the laser resonance chromatography (LRC) project \cite{laatiaoui2020a}, a great deal of attention became focused on developing element-selective spectroscopy that is conceptually dedicated to SHE ions. In the newly developed method, optical pumping of ions drifting in dilute helium is exploited to identify optical resonances. Successful excitation of ionic levels initiates pumping to metastable states and causes an abrupt change of the transport properties, which can be measured using drift time spectrometers \cite{laatiaoui2012,laatiaoui2020a}. However, to enable atomic structure investigations on systems that lack tabulated spectral lines, such experiments have to be pursued hand in hand with high-accuracy \textit{ab initio} calculations. \par 
Predicting energy levels, lifetimes, and branching ratios help scientists to quantify experimental parameters such as the required detector sensitivities and beam times \cite{laatiaoui2020b}. In a recent work \cite{kahl2019a}, the energy spectrum of the SHE ion Lr$^+$ (Z = 103) was predicted using the relativistic Fock space coupled cluster (FSCC) method and the configuration interaction approach combined  with many-body perturbation theory (CI + MBPT). The ground and the metastable excited states of Lr$^+$ stemming from the 7\textit{s}$^2$ and 6\textit{d}$^1$7\textit{s}$^1$ electron configurations, respectively, were identified together with the excitation scheme that is suitable for future LRC experiments \cite{laatiaoui2019a}. In this work, we are focusing our interest on the energy levels and the spectroscopic properties of the  SHE ion Rf$^+$ (Z = 104). The electronic structure of Rf is not trivial due to the complex combination of quantum interactions involving electron correlation, relativistic effects, and hyperfine structure \cite{indelicato2011,kaldor1998}. Nonetheless, Rf has been in the spotlight of many theoretical investigations. Since the early 90s, several studies have been devoted to predictions of its energy levels \cite{martin1996}, atomic radii \cite{johnson1990}, ionisation potentials \cite{johnson1990}, oxidation states, and chemical properties \cite{pershina1994,liu1999,anton2003,pershina2014}. For the Rf$^+$ ion, to the best of our knowledge, very few theoretical data are found. FSCC calculations revealed the ground and some excited states belonging to the 6\textit{d}$^1$7\textit{s}$^2$ and 7\textit{s}$^2$7\textit{p}$^1$ configurations \cite{eliav1995}. However, many levels were omitted, namely those originating from the metastable 6\textit{d}$^2$7\textit{s}$^1$ configuration, due to the practical restriction of the FSCC method to systems of up to two valence electrons or holes, which leaves part of the Rf$^+$ energy spectrum out of the scope of this approach. Therefore, here we are in particular interested in going beyond the earlier work \cite{eliav1995} by investigating the relative positions of the energy levels of the metastable 6\textit{d}$^2$7\textit{s}$^1$ configuration, which are important for the development of optical pumping schemes for the Rf$^+$ ion in future LRC experiments \cite{laatiaoui2020a,laatiaoui2020b}. \par 
The use of configuration interaction model prevails as the most appropriate method for treating the Rf$^+$ ion system, due to the fact that it can be applied to open shells that contain more than two valence particles and because of the straightforward approach to extracting the spectroscopic properties. In this work, the theoretical results are obtained using state-of-the-art relativistic approach \textit{via} the four-component Dirac-Coulomb Hartree-Fock (DCHF) calculation complemented with a multireference configuration interaction (MRCI) model as implemented in the DIRAC program package \cite{DIRAC19}. In MRCI, a subset of the full configuration interaction expansion is used to retrieve the correlation energy. In practice, only single and double excitations are retained up to the level of truncation of the configuration interaction expansion. MRCI models have been shown to yield accurate results for heavy and superheavy elements with many application available in the literature \cite{parmar2013,bross2015,kovacs2015}. We preferred the molecular DIRAC package \cite{DIRAC19} over the available  atomic codes \cite{desclaux1975,froesefischer2019,kahl2019b,fritzsche2019} as it can be also used to study the Rf$^+$-He interactions, which are very important for predicting transport properties of ions in gases \cite{visentin2020}. Such calculations will constitute the next step in our investigations of the properties of Rf$^+$. In order to assess the accuracy of the Rf$^+$ results, we also performed calculation of the properties of its lighter homologue, the Hf$^+$ ion, and compared our results with the experimental data that are systematically tabulated within the framework of the National Institute of Standards and Technology (NIST) spectral database \cite{nist2020}.

\section{Method and computational details}
All the calculations were carried out using the DIRAC19 code \cite{DIRAC19} and were based on the 4-component Dirac-Coulomb Hamiltonian. We used the finite-nucleus model in the form of a Gaussian charge distribution  to treat the nuclei \cite{VisDya97}. We have employed the Dyall basis sets for both elements \cite{dyall2004,dyall2011}. Preliminary tests carried out in our work showed that basis set expansion had a rather small effect on the energy levels (See also Supplementary Material Table S1). It is noteworthy, however, to point that orbitals with higher angular momentum contribute to the energies in the alkali atoms \cite{safronova1998}. Therefore, unless stated otherwise, all the results reported here were obtained using the dyall.cv3z basis sets, following the DIRAC19 nomenclature \cite{DIRAC19}. These basis sets consist of uncontracted Gaussian type-orbitals for the large component wavefunction up to (30\textit{s}24\textit{p}15\textit{d}11\textit{f}4\textit{g}1\textit{h}) and (32\textit{s}29\textit{p}20\textit{d}14\textit{f}4\textit{g}1\textit{h}) for the Hf and Rf elements \cite{dyall2004,dyall2011}, respectively. The small component functions are generated from the large component basis set by strict kinetic balance \cite{stanton1984}. We have also tested the singly and doubly augmented basis sets (s-aug-dyall.cv3z and d-aug-dyall.cv3z) by adding extra diffuse functions in an even-tempered manner; the augmentation also had only a small effects on the calculated energy levels (See Supplementary Material Table S2). \par 
The spherical symmetry of the atomic systems was reduced to the $D_{\infty h}$ symmetry group as is implemented in DIRAC19 \cite{DIRAC19}. In practice, the calculations use a subgroup of the axial rotation double group $D_{32h}$. In this double group, all $m_j$ values fall into unique representations as long as $m_j$ $\leq$ 32. Therefore, the Fock matrix was block-diagonalized in the orthonormal basis of the $m_j$ quantum numbers \cite{saue2020}. The atomic spinors were selectively discriminated with respect to the representation $\omega$, including the $m_j$ value and the parity. The atomic spinors that were used for the CI calculation (\textit{vide infra}) were obtained by using the average of configuration (AOC) type calculation at the DCHF level of theory \cite{slater1960}. The AOC allowed us to represent the open-shell system with 3 valence electrons that were evenly distributed over 12 valence spinors (6 Kramers pair) of \textit{s} and \textit{d} atomic character. Thus, 68 and 100 electrons were restricted to closed-shells for the Hf$^+$ and Rf$^+$ ions, respectively, whereas we used fractional occupation numbers (0.1250 = 3/12) for the merged Hf 6\textit{s}-5\textit{d} as well as Rf 7\textit{s}-6\textit{d} shells. \par 
The configuration interaction calculations were performed by using the Kramers-restricted configuration interaction (KRCI) module in DIRAC19 \cite{DIRAC19}. KRCI was developed from string-based configuration interaction \cite{thyssen2008,knecht2010}, the algorithm of which was fully operational within the 4-component relativistic framework \cite{saue2020}. In DIRAC19 \cite{DIRAC19}, the KRCI calculations use the concept of generalized active space (GAS) \cite{fleig2003}, which enables MRCI calculations with single and double electron excitations for different GAS set-ups \cite{saue2020}. The MRCI model a priory takes into consideration the dynamical correlation of the active electrons \cite{fleig2012}. \par 

\begin{table}[htp]
\centering
\begin{minipage}{12cm}
% \begin{table}[htp]
\centering
\caption{Specification of the generalized active space (GAS) scheme used for the calculations on Hf$^+$ and Rf$^+$ (see the Text for details).}
\label{table1}
\begin{tabular}{ l| c| c| c| c}
\hline\hline
GAS       & \multicolumn{2}{|c|}{Accumulated}  & Number of & Characters\footnote{For Hf$^+$ and Rf$^+$, n = 6 and 7, respectively} \\
Space     & \multicolumn{2}{|c|}{Electrons}    & Kramers   &            \\
          & Min\footnote{\textit{x}, \textit{y} and \textit{z} are variables that control the electron excitation process attributed to the selective GAS}           & Max                & pairs     &            \\
\hline
1         & 10-\textit{x} & 10  & 5 (5/0)      & (n-2)\textit{d} \\
2         & 18-\textit{y} & 18  & 4 (1/3)      & (n-1)\textit{s}, (n-1)\textit{p} \\
3         & 32-\textit{z} & 32  & 7 (0/7)      & (n-2)\textit{f} \\
4         & 32            & 35  & 9 (6/3)      & n\textit{s}, (n-1)\textit{d}, n\textit{p} \\
5         & 35            & 35  & 102 (52/50)  & Virtual \\
\hline\hline
\end{tabular}
% \end{table}
\end{minipage}
\end{table}

We report in \autoref{table1} the GAS set-up together with the technical specifications that were important in the MRCI calculation. For both Hf$^+$ and Rf$^+$ calculations, we placed within GAS 1, 2, and 3 the thirty-two highest-lying fully occupied spinors (sixteen Kramers pairs) that formed the basis of the following representations: $\omega$ = 1/2$_g$ (3 Kramers pairs), 3/2$_g$ (2 Kramers pairs), 5/2$_g$ (1), 1/2$_u$ (4), 3/2$_u$ (3), 5/2$_u$ (2), and 7/2$_u$ (1). These spinors are predominantly of \textit{d}, \textit{s}, \textit{p} and \textit{f} atomic characters. Furthermore, we placed within GAS 4 the twelve spinors with fractional electron occupation that form the basis of the representations $\omega$ = 1/2$_g$ (3 Kramers pairs), 3/2$_g$ (2) and 5/2$_g$ (1); as well as the six virtual spinors that form the basis of the representations $\omega$ = 1/2$_u$ (2 Kramers pairs) and 3/2$_u$ (1)). These spinors are predominantly of valence \textit{s}, \textit{d} and \textit{p} atomic characters. Finally, we placed within GAS 5 the 204 lowest-lying energy of virtual spinors (102 Kramers pairs) with energies below 30 atomic units. \par 
Within the defined GAS set up, the MRCI model was designed to activate in total 35 electrons, a method that we refer to as MRCI(35). We defined the parameters \textit{x}, \textit{y}, and \textit{z} to control the electron excitation process that occurred at the semi-core level of Hf$^+$ and Rf$^+$ (\autoref{table1}). These parameters took 0, 1 or 2 values, which signified zero-, single-, and double-electron excitations allowed from the selective GAS. We requested the following number of roots in the MRCI calculations: 37, 32, 22, 12, 6, and 1 roots for representations with even parity $\Omega$ = 1/2$_g$, 3/2$_g$, 5/2$_g$, 7/2$_g$, and 9/2$_g$, respectively; 25, 20, 12, 5, and 1 roots for representations with odd parity $\Omega$ =  1/2$_u$, 3/2$_u$, 5/2$_u$, 7/2$_u$, 9/2$_u$ and 11/2$_u$, respectively. The large number of roots were needed due to many near-degenerate electronic states that we found in the energy range between 0 to 50000 cm$^{-1}$ for both Hf$^+$ and Rf$^+$ ions.

\section{Results and Discussion}
\subsection{Energy Levels}
MRCI method is in practise limited by the choice of basis sets and the number of correlated electrons. It thus becomes important to investigate the effect of the basis set quality on the calculated energy levels of the Hf$^+$ ion in order: to pursue the basis set limit at a reasonable computational cost; and to ensure that uncertainties are maintained at acceptable levels. In the Supplementary Material (Table S1), we report the calculated energies as function of the basis sets quality. These multiplet energies include the ground state $^2$D$_{3/2}$ (5\textit{d}$^1$6\textit{s}$^2$) and the low-lying excited states $^2$D$_{5/2}$ (5\textit{d}$^1$6\textit{s}$^2$), $^4$F$_{3/2}$ (5\textit{d}$^2$6\textit{s}$^1$), and $^4$F$_{3/2}$ (5\textit{d}$^1$6\textit{s}$^1$6\textit{p}$^1$), which were selected to represent the whole manifold of the low lying  Hf$^+$ electronic levels. We used the uncontracted Dyall basis sets of double- (cv2z), triple- (cv3z), and quadruple-zeta (cv4z) zeta quality including valence-correlating and core-valence correlating functions \cite{dyall2004}. The quadrupole-zeta basis sets are also characterised by the presence of higher angular momentum up to $l = 6$ \textit{i} functions. By increasing the basis set quality (from double-zeta to quadruple-zeta), we only obtained small effect on the calculated energy levels, with the estimated standard deviations of the energies in the magnitude of tens of cm$^{-1}$ (see Table S1). Further augmentation by single and double diffuse functions at the triple-zeta basis set level is also found to have only a minor effect on the atomic energy levels, with the estimated standard deviations of the calculated energies ranging from less than 3 cm$^{-1}$ to 15 cm$^{-1}$ (see Supplementary Material, Table S2). Therefore, the uncertainties of the calculation due to basis set expansion and quality are small. We have thus selected the core-valence correlating triple-zeta basis set, also for consistency with earlier studies of analogous elements. For example, Fleig and Nayak \cite{fleig2013} reported similar MRCI calculations of the HfF$^+$ molecule using basis sets of triple-zeta quality. \par
We also investigated the effect of the semi-core-electron excitations on the energy levels of the Hf$^+$ ion. Although we are fundamentally interested in the ground and low-lying excited states that belong to the active space of the Hf 5\textit{d}, 6\textit{s} and 6\textit{p} orbitals, early configuration interaction studies recommended the consideration of core-electrons \cite{fleig2013}. To study this, we performed calculations in which the parameters \textit{x}, \textit{y} and \textit{z} in \autoref{table1} are varied. Since \textit{x}, \textit{y} and \textit{z} cannot be equal or higher than three (technically possible but generating a massive number of Slater determinants that are beyond the scope of our computational resources), we excluded this situation from the MRCI calculation. In Supplementary Material Table S3, we report the calculated multiplet energies as function of the \textit{x}, \textit{y}, and \textit{z} variables. We did not find major energy changes, when up to one electron excitation is allowed from GAS 1 (i.e. \textit{x} = 1). However, by allowing single and double electron excitation from GAS 2 (i.e. \textit{y} = 2), the energies of the states that have even parity were generally improved in relation to the experimental values. Moreover, we also found that the energies of the states with odd parity were shifted to higher values, with better agreement with the experiments, when one electron excitations were allowed in GAS 3 (i.e. \textit{z} = 1). Based on this study, MRCI(35) with \textit{x} = 0, \textit{y} = 2 and \textit{z} = 1 is our method of choice. This computational set-up explicitly treats correlation effects of 35 electrons, whereas the configuration interaction space contains the (5\textit{d},6\textit{s},6\textit{p})$^3$ configurations, with up to 2 holes in Hf 5\textit{s} and 5\textit{p} and 1 hole in Hf 4\textit{f}.\par
 
\begin{table}[htp]
\centering
\caption{Calculated energies (in cm$^{-1}$) of the low-lying excited states of the Hf$^+$ ion together with the recommended values (Final) that take into consideration the Breit ($\Delta_{B}$) and QED ($\Delta_{B+QED}$) corrections, classified with respect to the dominant electron configuration (Config.) and compared with the experimental values (Exp.).}
\label{table2}
\begin{tabular}{ l| c| c| c| c| c| c| c}
\hline\hline
Config.                                       & State &  J  & Exp.& \multicolumn{4}{c}{Theory}  \\
                                              &       &     &     & MRCI&$\Delta_{B}$&$\Delta_{B+QED}$&Final   \\
\hline
5\textit{d}$^1$6\textit{s}$^2$                & $^2$D & 3/2 &0    &0    &-   &-   &    0\\
                                              &       & 5/2 &3051 &2850 &-67 &-5  & 2845\\
5\textit{d}$^2$6\textit{s}$^1$                & $^4$F & 3/2 &3642 &3970 &101 &-95 & 3875\\
                                              &       & 5/2 &4905 &4913 &63  &-92 & 4821\\
                                              &       & 7/2 &6344 &6165 &19  &-93 & 6072\\
                                              &       & 9/2 &8362 &7835 &-43 &-90 & 7745\\
                                              & $^4$P & 1/2 &11952&11745&46  &-96 &11649\\
                                              &       & 3/2 &12921&12976&11  &-93 &12883\\
                                              &       & 5/2 &13486&13396&0.1 &-91 &13305\\
                                              & $^2$F & 5/2 &12071&12227&65  &-100&12127\\
                                              &       & 7/2 &15085&14480&-16 &-100&14380\\
                                              & $^2$D & 3/2 &14360&14430&34  &-106&14324\\
                                              &       & 5/2 &17369&16834&-74 &-84 &16750\\
                                              & $^2$P & 1/2 &15255&15366&37  &-133&15233\\
                                              &       & 3/2 &17830&17945&-56 &-118&17827\\
                                              & $^2$G & 9/2 &17389&17394&-24 &-106&17288\\
                                              &       & 7/2 &17711&18100&-24 &-108&17992\\
                                              & $^2$S & 1/2 &  -  &21117&-87 &-95 &21022\\
5\textit{d}$^3$                               & $^4$F & 3/2 &18898&19605&138 &-221&19384\\
                                              &       & 5/2 &20135&20560& 95 &-220&20340\\
                                              &       & 7/2 &21638&21745& 44 &-217&21528\\
                                              &       & 9/2 &23146&23023&-6  &-215&22808\\
                                              & $^4$P & 1/2 &26997&27689&61  &-218&27471\\
                                              &       & 3/2 &27285&27930&56  &-217&27713\\
                                              &       & 5/2 &28547&28933&12  &-217&28716\\
5\textit{d}$^1$6\textit{s}$^1$6\textit{p}$^1$ & $^4$F & 3/2 &28069&27320&9   &-89 &27231\\
                                              &       & 5/2 &29405&28666&-1  &-97 &28569\\
                                              &       & 7/2 &33776&32484&-97 &-82 &32402\\
                                              &       & 9/2 &38186&36836&-161&-78 &36758\\
                                              & $^4$D & 1/2 &29160&28713&-50 &-56 &28657\\
                                              &       & 3/2 &31784&31214&-97 &-81 &31133\\
                                              &       & 5/2 &34355&33549&-63 &-90 &33459\\
                                              &       & 7/2 &36882&35578&-100&-84 &35494\\
                                              & $^2$D & 5/2 &33181&32390&-75 &-100&32290\\
                                              &       & 3/2 &34124&33174&-17 &-98 &33076\\
                                              & $^2$P & 1/2 &33136&32693&-41 &-145&32548\\
                                              &       & 3/2 &36373&35973&-82 &-139&35834\\
                                              & $^2$D & 3/2 &37886&38237&-92 &-187&38050\\
                                              &       & 5/2 &41761&41313&-121&-172&41141\\
                                              & $^2$F & 5/2 &38579&37873&-54 &-126&37747\\
                                              &       & 7/2 &41407&40840&-97 &-125&40715\\
                                              & $^4$P & 1/2 &38399&38271&-56 &-82 &38189\\
                                              &       & 3/2 &39227&38546&-100&-70 &38476\\
                                              &       & 5/2 &40507&39585&-129&-107&39478\\
5\textit{d}$^2$6\textit{p}$^1$                & $^4$G & 5/2 &34943&34585&10  &-168&34417\\
                                              &       & 7/2 &38499&37751&-29 &-195&37556\\
                                              
\hline\hline
\end{tabular}
\end{table}

We report in \autoref{table2} the calculated excitation energies of Hf$^+$ obtained using the MRCI(35) model together with known experimental values for comparison \cite{sansonetti2005}. Only levels with energies below 40000 cm$^{-1}$ are listed for convenience. We have analysed the natural orbital occupation numbers of the configuration interaction vectors to classify the states and to identify their dominant electron configurations. 
In order to correct our results for the Breit and the lowest order QED contributions \cite{indelicato2007}, we used the GRASP program package \cite{froesefischer2019}, which is based on the Dirac-Coulomb-Breit Hamiltonian and the multi-configuration Dirac-Fock (MCDF) model and also incorporates the lowest-order QED corrections, i.e. the vaccuum polarization and the self-energy terms. Details for the implementation of the Breit and QED corrections in GRASP\cite{froesefischer2019} can be found elsewhere \cite{dyall1989,jonsson2007,froesefischer2019}. The reference space for the MCDF configuration interaction calculation was the 4\textit{f}$^{14}$(5\textit{d}6\textit{s}6\textit{p})$^3$ multiplet manifold. The core 5\textit{s} and 5\textit{p} electrons were also correlated, and they produced an average shift in the transition energies of only a few cm$^{-1}$. The virtual space for the configuration interaction expansion consisted in one extra spinor for each \textit{l} quantum number from 0 to 4 (i.e. 7\textit{s}7\textit{p}6\textit{d}5\textit{f}5\textit{g}). It was also found that the calculated Breit and QED corrections were relatively independent of the size of the CI expansion and they remained very close to the values listed in \autoref{table2}. The values in \autoref{table2} represent the differences in energy calculated at the Dirac-Coulomb and Dirac-Coulomb-Breit level of theory ($\Delta_{B}$) and the difference between the Dirac-Coulomb-Breit and Dirac-Coulomb-Breit + QED values ($\Delta_{B+QED}$). We found that the Breit and QED corrections are relatively small, on the order of 100 cm$^{-1}$ for Hf$^+$, and comparable with the Breit and QED effects calculated for analogous 5\textit{d} elements \cite{indelicato2007,pasteka2017,kahl2019a}. Based on the calculated MRCI energy data together with the Breit and QED corrections, we obtain the recommended energy values for Hf$^+$ that are also listed in \autoref{table2}. \par
The Hf$^+$ ground and low lying excited states belong to the configurations 5\textit{d}$^1$6\textit{s}$^2$, 5\textit{d}$^2$6\textit{s}$^1$, 5\textit{d}$^3$ that have even parity, as well as configurations 5\textit{d}$^1$6\textit{s}$^1$6\textit{p}$^1$ and 5\textit{d}$^2$6\textit{p}$^1$ that have odd party. We also find that the multiplets corresponding to the 6\textit{s}$^2$6\textit{p}$^1$  configuration are definitely above 50000 cm$^{-1}$. The ground state is four-fold degenerate with J = 3/2 that arise from the spin-orbit coupling of the $^2$D term of configuration 5\textit{d}$^1$6\textit{s}$^2$. The mean percentage error for the states that belong to the 5\textit{d}$^2$6\textit{s}$^1$ configuration is relatively small ($<$5$\%$). We found larger errors for the states originating in the configuration where the 3 electrons occupy different orbitals (5\textit{d}$^1$6\textit{s}$^1$6\textit{p}$^1$). Overall, the calculated energy levels of Hf$^+$ are in good agreement with the NIST values\cite{sansonetti2005}, confirming the suitability of the MRCI model for this study.\par
However, we should point out that the calculated spin-orbit splitting of the $^2$D ground state, which is off by 200 cm$^{-1}$, is somewhat intriguing (see \autoref{table2}). We have tested the possibility that the error stems from the use of the AOC procedure in the DCHF calculation (Method section). We carried out an additional test by changing the fractional occupation scheme of the Hf 6\textit{s} and 5\textit{d} orbitals, the AOC itself being necessary to populate unpaired electrons on degenerate spinors. We eventually found out that by placing 2 electrons in 6\textit{s} (in other words, taking 6\textit{s} as a closed shell) and 1 electron in 5\textit{d}$_{3/2}$ (in a fractional manner), the spin-orbit splitting of the Hf$^+$ $^2$D ground state equalled 2910 cm$^{-1}$, a value which was very close to the experimental data. However, all the energy levels of the 5\textit{d}$^2$6\textit{s}$^1$ configuration were shifted to higher energy values as a side effect. We thus proceeded with the original AOC scheme.\par

\begin{table}[htp]
\centering
\caption{Calculated energies (in cm$^{-1}$) of the low-lying excited states of Rf$^+$ ion together with the recommended values (Final) that take into consideration the Breit ($\Delta_{B}$) and QED ($\Delta_{B+QED}$) corrections, classified with respect to the dominant electron configuration (Config.).}
\label{table3}
\begin{tabular}{ l| c| c| c| c| c| c}
\hline\hline
Config.                                       & State &  J  & \multicolumn{4}{c}{Theory} \\
                                              &       &     & MRCI   & $\Delta_{B}$ & $\Delta_{B+QED}$ & Final \\
\hline
6\textit{d}$^1$7\textit{s}$^2$                & $^2$D & 3/2 &0    &-&  - &0\\
                                              &       & 5/2 &5682 &-177&-2  &5680\\
6\textit{d}$^2$7\textit{s}$^1$                & $^4$F & 3/2 &15931&94  &-253&15678\\
                                              &       & 5/2 &17642&42  &-250&17392\\
                                              &       & 7/2 &20476&-71&-245&20231\\
                                              &       & 9/2 &23621&-172&-239&23382\\
                                              & $^4$P & 1/2 &24864&13  &-249&24615\\
                                              &       & 3/2 &26993&-35 &-245&26648\\
                                              &       & 5/2 &29832&-74 &-245&29587\\
                                              & $^2$F & 5/2 &26820&-69 &-255&26565\\
                                              &       & 7/2 &32631&-225&-255&32376\\
                                              & $^2$D & 3/2 &30245&-77 &-262&29983\\
                                              &       & 5/2 &34515&-244&-236&34279\\
                                              & $^2$P & 1/2 &32851&-49 &-301&32550\\
                                              &       & 3/2 &36860&-210&-290&36570\\
                                              & $^2$G & 9/2 &34267&-156&-257&34010\\
                                              &       & 7/2 &36615&-98&-254&36361\\
                                              & $^2$S & 1/2 &44529&-261&-233&44296\\
7\textit{s}$^2$7\textit{p}$^1$                & $^2$P & 1/2 &16691&-122&-34 &16657\\
                                              &       & 3/2 &31288&-223&-47 &31241\\
6\textit{d}$^1$7\textit{s}$^1$7\textit{p}$^1$ & $^4$F & 3/2 &28052&-44 &-206&27846\\
                                              &       & 5/2 &31244&-77 &-213&31031\\
                                              &       & 7/2 &38140&-252&-197&37943\\
                                              &       & 9/2 &50467&-399&-199&50268\\
                                              & $^4$D & 1/2 &36338&-87 &-182&36156\\
                                              &       & 3/2 &39040&-192&-226&38814\\
                                              &       & 5/2 &42676&-169&-266&42410\\
                                              &       & 7/2 &47934&-248&-197&47737\\
                                              & $^2$D & 5/2 &37601&-220&-203&37398\\
                                              &       & 3/2 &42421&-227&-226&42195\\
                                              & $^2$F & 5/2 &46318&-195&-266&46052\\
                                              &       & 7/2 &55434&-252&-379&55055\\
                                              & $^2$D & 3/2 &48008&-160&-204&47804\\
                                              &       & 5/2 &53780&-180&-279&53501\\
                                              & $^4$P & 1/2 &48374&-193&-210&48164\\
                                              &       & 3/2 &50577&-193&-314&50263\\
                                              &       & 5/2 &51921&-300&-345&51576\\
                                              
\hline\hline
\end{tabular}
\end{table}

We report in \autoref{table3} the excitation energies of Rf$^+$ calculated by means of the MRCI(35) methodology. For convenience, only levels with energies below 50000 cm$^{-1}$ are listed. The spin-orbit coupling interactions of the 6\textit{d}, 7\textit{s} and 7\textit{p} electrons are much larger than for Hf$^+$ and the energy spectrum is rather more sparse. We have also performed a similar analysis to account for the Breit and the QED contributions as was done for the Hf$^+$ ion using MCDF calculation in GRASP\cite{froesefischer2019}. The reference space for the MCDF configuration interaction calculation was the 5\textit{f}$^{14}$(6\textit{d}7\textit{s}7\textit{p})$^3$ multiplet manifold. The core 6\textit{s} and 6\textit{p} electrons were also correlated. The virtual space for the configuration interaction expansion consisted in one extra spinor for each \textit{l} quantum number from 0 to 2, together with two extra spinors for each \textit{l} quantum number from 3 to 5 and 6\textit{h} function (i.e. 8\textit{s}8\textit{p}7\textit{d}6\textit{f}7\textit{f}5\textit{g}6\textit{g}6\textit{h}). The Breit and QED contributions approximately double going from Hf$^+$ to Rf$^+$ ions and are on the order of 200 cm$^{-1}$ (see \autoref{table3}) for most levels, in good agreement with calculated QED effects for the analogous SHE elements \cite{fritzsche2005,indelicato2007,fritzsche2007,kahl2019a}. The recommended energy values for Rf$^+$ are also listed in \autoref{table3} based of the calculated MRCI energy data together with the Breit and QED corrections. \par
We identify the multiplet terms from configurations 6\textit{d}$^1$7\textit{s}$^2$, 6\textit{d}$^2$7\textit{s}$^1$, and  7\textit{s}$^2$7\textit{p}$^1$ as the ground and low-lying excited states. In particular, the two spin-orbit components of the $^2$P state (7\textit{s}$^2$7\textit{p}$^1$) are predicted to reside at a much lower energy than in Hf$^+$. Similar to Hf$^+$, the calculated ground state of Rf$^+$ is $^2$D$_{3/2}$ from the  6\textit{d}$^1$7\textit{s}$^2$ configuration and the first low-lying excited state is its spin-orbit counterpart ($^2$D$_{5/2}$) at 5682 cm$^{-1}$. The second low-lying excited state is the metastable state ($^4$F$_{3/2}$) that forms the basis of the 6\textit{d}$^2$7\textit{s}$^1$ configuration. Above this level, we obtain the third excited state that is a term with odd parity ($^2$P$_{1/2}$ from configuration 7\textit{s}$^2$7\textit{p}$^1$). The energy difference between this latter and the metastable state is predicted to be 761 cm$^{-1}$. \par 
The definition of the second and third excited states in the Rf$^+$ energy spectrum is critical in the set-up of the ion-mobility-assisted laser spectroscopy studies (\textit{vide infra}). In particular, the relative position of the $^2$P$_{1/2}$ odd level  with respect to the $^4$F$_{3/2}$ metastable state could determine the feasibility of one of the proposed pumping schemes, see below. To evaluate the energy uncertainty for these levels from a theoretical perspective, we used the Fock space coupled cluster (FSCC) method \cite{eliav1994a,eliav1994b}. However, we note once again that FSCC is currently not applicable for all the states of interest in Rf$^+$, and we thus focus on those states that can be reached with this method. The FSCC calculation was also carried out using the DIRAC19 code, and the computational details were kept, inasmuch as possible, similar to the present MRCI method for consistency. We started with a relativistic DCHF calculation of the Rf$^{2+}$ ion because the valence electron operator in the FSCC is defined with respect to a closed-shell reference \cite{eliav1994a,eliav1994b}. Then all the closed-shell electrons were correlated and virtual orbitals were also included up to +30 atomic units. Finally, one electron was added in order to obtain Rf$^+$, and the coupled cluster equations were solved accordingly. Within the FSCC nomenclature, we have applied sector(0,1) with respect to the closed-shell reference, where 0 signifies 0 holes in the 7\textit{s} valence orbital and 1 signifies 1 valence electron that is allowed to occupy the Rf 6\textit{d} and 7\textit{p} orbitals. Thus only excitation energies that result from the 6\textit{d}$^1$7\textit{s}$^2$ and 7\textit{s}$^2$7\textit{p}$^1$ configurations could be evaluated. \par 
The calculated FSCC energies of the spin-orbit components of the $^2$P terms (configuration 7\textit{s}$^2$7\textit{p}$^1$) are equal to 17535 and 33785 cm$^{-1}$ when using the triple-zeta basis set. The same energies became 18216 and 34502 cm$^{-1}$ upon switching to the quadruple-zeta basis set. These levels thus present basis set dependency but they remained definitely at a higher energy than the MRCI $^2$P results listed in \autoref{table3}. For the $^2$D$_{5/2}$ level (configuration 6\textit{d}$^1$7\textit{s}$^2$), the FSCC energy equalled 7064 and 7334 cm$^{-1}$, respectively by using the triple- and quadruple-zeta basis sets; also significantly higher than the MRCI predictions. In another FSCC study of Rf$^+$, dated back to 1995 \cite{eliav1995}, the authors have used a very large basis set but they have considered only correlation of 34 external electrons. Furthermore, they studied electron correlation effects by varying the size of the active space of virtual orbitals that were included in their calculation. The reported energy of the $^2$P$_{1/2}$ state is within the 17300 - 19400 cm$^{-1}$ range, whereas the $^2$P$_{3/2}$ is within 34290 - 35381 cm$^{-1}$, and the $^2$D$_{5/2}$ is calculated at approximately 7300 cm$^{-1}$, in good agreement with the present FSCC calculations. From the estimated standard deviation of the energies that are obtained using the MRCI and the FSCC methods, we can derive the uncertainties of the $^2$D$_{5/2}$, $^2$P$_{1/2}$ and $^2$P$_{3/2}$ states as follows (in cm$^{-1}$): 5682 $\pm$ 1382, 16691 $\pm$ 844 and 31283 $\pm$ 2502, respectively. \par 
What is striking, though, is that the calculated uncertainties are relatively large for the states that are mainly driven by the spin-orbit coupling interaction. The FSCC energy separation between the $^2$P$_{1/2}$ and $^2$P$_{3/2}$ states is 10 $\%$ larger than that reported in \autoref{table3} (we obtain 14600 cm$^{-1}$ (MRCI) versus 16250 cm$^{-1}$ (FSCC)). Similarly, the energy separation between the $^2$D$_{3/2}$ and the $^2$D$_{5/2}$ states is found 25$\%$ higher for FSCC compared to the MRCI results reported in \autoref{table3} (7064 cm$^{-1}$ (FSCC) vs. 5682 cm$^{-1}$ (MRCI)). It is possible that the \textit{modus operandi} of the FSCC calculation produces overestimation of the energies, due to the fact that it does not account for the mixing with the 3-valence electron configurations. 

\subsection{Spectroscopic Properties}
The electronic states that originate from the configurations (n-1)\textit{d}$^1$n\textit{s}$^1$n\textit{p}$^1$ and n\textit{s}$^2$n\textit{p}$^1$ decay \textit{via} electric-dipole E1 mechanism for both Hf$^+$ (with n = 6) and Rf$^+$ (with n = 7) ions. However, the electronic states that originate from the configurations (n-1)\textit{d}$^1$n\textit{s}$^2$ and (n-1)\textit{d}$^2$n\textit{s}$^1$ decay only \textit{via} electric-quadrupole E2 and magnetic-dipole M1 mechanisms. We calculated the inter- and intra-configuration transition probabilities by using a phenomenological effective Hamiltonian. By means of the \textit{ab initio} MRCI method, the relativistic form of the transition moment operator was also used to derive transition probabilities at the E1 level.
But, we did not obtain the transition probabilities of E2 and M1 transitions from \textit{ab initio} calculations; instead, we have turned to the effective Hamiltonian approach. It is noteworthy that the effective Hamiltonian method constitutes a semi-empirical approach for spectroscopic properties. The results are based upon mathematical least square fit of the \textit{ab initio} energy levels, and thereby produce qualitative transition probabilities that are sufficient to develop optical pumping scenarios for the Rf$^+$ ion (see section III C).
We use the quantum theory of Slater \cite{slater1960}, where the matrix elements of the effective Hamiltonian are built based on perturbation theory and the central field approximation.  \par
Slater's theory is described in detail in many texts \cite{slater1960,judd1963,cowan1981}, and we give here only a brief overview. The energies and the eigenvectors of any spectroscopic states of a system with \textit{N} electrons are generally obtained from the diagonalization of the matrix elements of the atomic Dirac-Coulomb Hamiltonian (in atomic units):
\begin{dmath}
H = \displaystyle\sum\limits_{i}^{N}h_{D}(i)+\displaystyle\sum\limits_{i<j}^{N}(\dfrac{1}{r_{ij}}),
\label{eq1}
\end{dmath}
where $h_D$ and $1/r_{ij}$ express the one- and two-electron operators, respectively. The one-electron Dirac operator consists of the kinetic energy and electron-nuclei attraction terms. The two-electron operator, on the other hand, includes the Coulomb-repulsion (\textit{J}) and the exchange (\textit{K}) integrals between electrons \textit{i} and \textit{j}. In Slater's theory for atomic calculations, $1/r_{ij}$ is expanded with respect to spherical harmonics so that the \textit{J} and \textit{K} integrals can be separately discriminated into parts according to the spin, angular, and radial components. We use this definition to construct our effective Hamiltonian ($H_{eff}$), and therefore transform Equation \ref{eq1} as follows:
\begin{dmath}
H_{eff} = \displaystyle\sum\limits_{i}^{N}h_{0}(i)+\displaystyle\sum\limits_{i}^{N}\zeta_{i}\; l_{i} \cdot s_{i}\\+\displaystyle\sum\limits_{i<j}^{N}\displaystyle\sum\limits_{k}(F^k(n_il_i,n_jl_j)f_k(l_im_{li},l_jm_{lj})\\-\delta_k(m_{si},m_{sj})G^k(n_il_i,n_jl_j)g_k(l_im_{li},l_jm_{lj})),
\label{eq2}
\end{dmath}
where the first and the second terms on the right-hand side of the equation represent the one-electron operators including the spin-orbit coupling \cite{misetich1964}, while the third term represents the two-electron operator. In this latter, the terms $f_k$, $g_k$ and $\delta_k$ (with \textit{k} being the multipole index) arise from the integration over the spin and angular components of the wavefunctions; the terms $F_k$ and $G_k$ are conventionally referred as the Slater-Condon integrals and result from the integration over the radial component \cite{cowan1981}. \par 
By using Equation \ref{eq2}, the effective Hamiltonian can be simply parameterized by the Slater-Condon integrals ($F_k$ and $G_k$) and spin-orbit coupling constants ($\zeta_{i}$) \cite{cowan1981}, which allows us to operate the configuration interaction algorithm in a semi-empirical manner. We note however that the parameters can be numerically evaluated, and there are numerous examples in the literature for dealing with lanthanides \cite{cowan1981,ramanantoanina2018,poe2021} or heavier actinide elements \cite{ramanantoanina2017,Ramanantoanina2019,albrecht-schmitt2020}. For the Hf$^+$ and Rf$^+$ ions, we constructed the effective Hamiltonian with \textit{N} = 3 electrons in the \textit{s}, \textit{d} and \textit{p} valence orbitals. In this case, the size of the matrix elements of the effective Hamiltonian in Equation \ref{eq2} equalled 816 $\times$ 816, which was the dimension of the Hilbert space spanned by the configuration interaction problem of 3 electrons in 18 spinors. Moreover, the summation over the multipole index \textit{k} did not exceed 4 in Equation \ref{eq2} \cite{cowan1981}. Then, the Slater-Condon integrals and spin-orbit coupling constants are calculated by least-square fit methods by minimizing the residual between the theoretical energies listed in \autoref{table2} and \autoref{table3} (our reference energy values) and the calculated energies from the effective Hamiltonian. The least-square fit is implemented using the fminsearch tool in Matlab \cite{MATLAB2015}. We report in the Supplementary Material Table S4 and Table S5 the calculated energies obtained from the effective Hamiltonian and compared with the reference values. The mathematical fit is well suited for well-separated energy levels; thus the discrepancies between the reference energies and the effective Hamiltonian are larger for Hf$^+$ than Rf$^+$ (see Table S4 and Table S5). 

We obtain the oscillator strengths using the following equations (in atomic units):
\begin{dmath}
f^{(E1)}_{i,j}=\dfrac{2}{3}(E(j)-E(i))\displaystyle\sum\limits_{\alpha}|\bra{\psi_j}D_\alpha\ket{\psi_i}|^2
\label{eq3}
\end{dmath}
\begin{dmath}
f^{(M1)}_{i,j}=\dfrac{2}{3}\alpha^2(E(j)-E(i))\displaystyle\sum\limits_{\alpha}|\bra{\psi_j}M_\alpha\ket{\psi_i}|^2
\label{eq4}
\end{dmath}
\begin{dmath}
f^{(E2)}_{i,j}=\dfrac{1}{20}\alpha^2(E(j)-E(i))^3\displaystyle\sum\limits_{a,b}|\bra{\psi_j}Q_{ab}\ket{\psi_i}|^2
\label{eq5}
\end{dmath}
where \textit{$D_\alpha$}, $Q_{ab}$ and \textit{$M_\alpha$} are the electric dipole moment operator and the electric quadrupole tensor, which are formulated in the length gauge, and the magnetic dipole moment operator, respectively; the terms within the bra-ket notations represent the transition probabilities between states \textit{i} and \textit{j}, referring to the lower and upper levels of the electronic transitions. \textit{E} and $\psi$ are the calculated energies and eigenfunctions of the effective Hamiltonian. In Equation \ref{eq4} and Equation \ref{eq5}, $\alpha$ = 1/137 denotes the fine structure constant. We derived the Einstein coefficients $\textit{A}_{E1}$, $\textit{A}_{M1}$ and $\textit{A}_{E2}$ from the calculated transition probabilities using Equation S1, Equation S2 and Equation S3 in Supplementary Material, respectively.

\begin{table}[htp]
\centering
\caption{Calculated Einstein coefficients ($\textit{A}_{E1}$, [s$^{-1}$]) and  branching ratios ($\beta$) for the Hf$^+$ ion, compared with the experimental data.} %($\textit{A}_{E1}$ exp. in [s$^{-1}$])}
\label{table4}
\begin{tabular}{ l c| c c| c| c| c}
\hline\hline
\multicolumn{2}{c|}{Upper Level} & \multicolumn{2}{|c|}{Lower Level} & $\textit{A}_{E1}$ Exp. & $\textit{A}_{E1}$ & $\beta$ \\
\hline
5\textit{d}$^1$6\textit{s}$^1$6\textit{p}$^1$ & $^4$F$_{3/2}$ & 5\textit{d}$^1$6\textit{s}$^2$ & $^2$D$_{3/2}$ &            &1.271 10$^7$&0.520 \\
                                              &               & 5\textit{d}$^1$6\textit{s}$^2$ & $^2$D$_{5/2}$ &            &1.828 10$^4$&0.000 \\
                                              &               & 5\textit{d}$^2$6\textit{s}$^1$ & $^4$F$_{3/2}$ &            &1.068 10$^7$&0.437 \\
                                              &               & 5\textit{d}$^2$6\textit{s}$^1$ & $^4$F$_{5/2}$ &            &5.902 10$^2$&0.000 \\
5\textit{d}$^1$6\textit{s}$^1$6\textit{p}$^1$ & $^4$F$_{5/2}$ & 5\textit{d}$^1$6\textit{s}$^2$ & $^2$D$_{3/2}$ &3.100 10$^7$&1.953 10$^7$&0.552 \\
                                              &               & 5\textit{d}$^1$6\textit{s}$^2$ & $^2$D$_{5/2}$ &            &4.894 10$^6$&0.138 \\
                                              &               & 5\textit{d}$^2$6\textit{s}$^1$ & $^4$F$_{3/2}$ &            &3.455 10$^6$&0.098 \\
                                              &               & 5\textit{d}$^2$6\textit{s}$^1$ & $^4$F$_{5/2}$ &            &5.362 10$^6$&0.152 \\
                                              &               & 5\textit{d}$^2$6\textit{s}$^1$ & $^4$F$_{7/2}$ &            &1.561 10$^5$&0.004 \\
5\textit{d}$^1$6\textit{s}$^1$6\textit{p}$^1$ & $^4$F$_{7/2}$ & 5\textit{d}$^1$6\textit{s}$^2$ & $^2$D$_{5/2}$ &2.100 10$^7$&7.554 10$^6$&0.246 \\
                                              &               & 5\textit{d}$^2$6\textit{s}$^1$ & $^4$F$_{5/2}$ &            &3.920 10$^6$&0.128 \\    
                                              &               & 5\textit{d}$^2$6\textit{s}$^1$ & $^4$F$_{7/2}$ &            &1.674 10$^7$&0.545 \\
                                              &               & 5\textit{d}$^2$6\textit{s}$^1$ & $^4$F$_{9/2}$ &            &1.518 10$^6$&0.049 \\
\hline\hline
\end{tabular}
\end{table}    

\begin{table}[htp]
\centering
\caption{Calculated Einstein coefficients ($\textit{A}_{E1}$, [s$^{-1}$]) and  branching ratios ($\beta$) for the Rf$^+$ ion.}
\label{table5}
\begin{tabular}{ l c| c c| c| c}
\hline\hline
\multicolumn{2}{c|}{Upper Level} & \multicolumn{2}{|c|}{Lower Level} & $\textit{A}_{E1}$ & $\beta$ \\
\hline
7\textit{s}$^2$7\textit{p}$^1$                & $^2$P$_{1/2}$ & 6\textit{d}$^1$7\textit{s}$^2$ & $^2$D$_{3/2}$ &1.089 10$^8$&0.977   \\
                                              &               & 6\textit{d}$^2$7\textit{s}$^1$ & $^4$F$_{3/2}$ &2.530 10$^6$&0.023    \\
6\textit{d}$^1$7\textit{s}$^1$7\textit{p}$^1$ & $^4$F$_{3/2}$ & 6\textit{d}$^1$7\textit{s}$^2$ & $^2$D$_{3/2}$ &1.633 10$^8$&0.783    \\
                                              &               & 6\textit{d}$^1$7\textit{s}$^2$ & $^2$D$_{5/2}$ &7.404 10$^6$&0.036    \\
                                              &               & 6\textit{d}$^2$7\textit{s}$^1$ & $^4$F$_{3/2}$ &3.575 10$^7$&0.171    \\
                                              &               & 6\textit{d}$^2$7\textit{s}$^1$ & $^4$F$_{5/2}$ &1.965 10$^6$&0.009    \\
                                              &               & 6\textit{d}$^2$7\textit{s}$^1$ & $^4$P$_{1/2}$ &8.039 10$^4$&$<$0.001 \\
                                              &               & 6\textit{d}$^2$7\textit{s}$^1$ & $^2$F$_{5/2}$ &1.026 10$^5$&$<$0.001 \\
                                              &               & 6\textit{d}$^2$7\textit{s}$^1$ & $^4$P$_{3/2}$ &9.612 10$^4$&$<$0.001 \\
6\textit{d}$^1$7\textit{s}$^1$7\textit{p}$^1$ & $^4$F$_{5/2}$ & 6\textit{d}$^1$7\textit{s}$^2$ & $^2$D$_{3/2}$ &2.415 10$^8$&0.535    \\
                                              &               & 6\textit{d}$^1$7\textit{s}$^2$ & $^2$D$_{5/2}$ &1.212 10$^8$&0.268    \\
                                              &               & 6\textit{d}$^2$7\textit{s}$^1$ & $^4$F$_{3/2}$ &2.585 10$^7$&0.057    \\
                                              &               & 6\textit{d}$^2$7\textit{s}$^1$ & $^4$F$_{5/2}$ &5.928 10$^7$&0.131    \\
                                              &               & 6\textit{d}$^2$7\textit{s}$^1$ & $^4$F$_{7/2}$ &2.394 10$^6$&0.005    \\
                                              &               & 6\textit{d}$^2$7\textit{s}$^1$ & $^2$F$_{5/2}$ &1.312 10$^6$&0.003    \\
                                              &               & 6\textit{d}$^2$7\textit{s}$^1$ & $^4$P$_{3/2}$ &5.521 10$^4$&$<$0.001 \\
                                              &               & 6\textit{d}$^2$7\textit{s}$^1$ & $^4$P$_{5/2}$ &5.504 10$^4$&$<$0.001 \\
                                              &               & 6\textit{d}$^2$7\textit{s}$^1$ & $^2$D$_{3/2}$ &1.653 10$^5$&$<$0.001 \\
\hline\hline
\end{tabular}
\end{table}  

We report in \autoref{table4} and \autoref{table5} the calculated Einstein coefficients and the branching ratios of the E1 atomic radiative transitions of the Hf$^+$ and Rf$^+$ ions, respectively. 
For clarity, we list only transitions that have potential implication in the optical pumping process of the LRC experiment \cite{laatiaoui2019a}. The levels include the spin-orbit manifolds of $^2$D ((n-1)\textit{d}$^1$n\textit{s}$^2$, with n = 6 and 7 for Hf$^+$ and Rf$^+$, respectively), as well as the low-lying metastable states $^4$F ((n-1)\textit{d}$^2$n\textit{s}$^1$) and the low-lying bright excited states $^4$F ((n-1)\textit{d}$^1$n\textit{s}$^1$n\textit{p}$^1$) and $^2$P (7\textit{s}$^2$7\textit{p}$^1$).
For the Hf$^+$ ion, the only two experimental values that are available from the literature are added in \autoref{table4} for comparison \cite{sansonetti2005}. We found that the calculated Einstein coefficients are underestimated when compared to the experimental values, but the relative strengths of the electric-dipole transitions are well reproduced. The strongest transitions are $^2$D$_{3/2}$ $\rightarrow$ $^4$F$_{5/2}$ for both Hf$^+$ and Rf$^+$ ions. In order to get an additional check on the presented predictions, the electric-dipole transition probabilities were also computed using the ab initio MRCI scheme in the Dirac19 program package \cite{DIRAC19}. In the Supplementary Material, Figure S1 and Figure S2 show the simulated absorption spectra of both Hf$^+$ and Rf$^+$ ions, respectively, as they are determined with the \textit{ab initio} and the effective Hamiltonian methods, revealing the consistency between the two theoretical models.\par 

\begin{table}[htp]
\centering
\caption{Calculated Einstein coefficients ($\textit{A}_{M1}$ and $\textit{A}_{E2}$, [s$^{-1}$]) for the magnetic-dipole (M1) and electric-quadrupole (E2) transitions, respectively, of the Rf$^+$ ion.}
\label{table6}
\begin{tabular}{ l c| c c| c| c}
\hline\hline
\multicolumn{2}{c|}{Upper Level} & \multicolumn{2}{|c|}{Lower Level} & $\textit{A}_{M1}$ & $\textit{A}_{E2}$ \\
\hline
6\textit{d}$^1$7\textit{s}$^2$ & $^2$D$_{5/2}$ & 6\textit{d}$^1$7\textit{s}$^2$ & $^2$D$_{3/2}$ &1.480          &1.445 10$^{-3}$\\
6\textit{d}$^2$7\textit{s}$^1$ & $^4$F$_{3/2}$ & 6\textit{d}$^1$7\textit{s}$^2$ & $^2$D$_{3/2}$ &3.165 10$^{-2}$&1.833 10$^{-2}$\\
                               &               & 6\textit{d}$^1$7\textit{s}$^2$ & $^2$D$_{5/2}$ &5.159 10$^{-2}$&3.051 10$^{-6}$\\
6\textit{d}$^2$7\textit{s}$^1$ & $^4$F$_{5/2}$ & 6\textit{d}$^1$7\textit{s}$^2$ & $^2$D$_{3/2}$ &5.001 10$^{-4}$&1.714 10$^{-1}$\\ 
                               &               & 6\textit{d}$^1$7\textit{s}$^2$ & $^2$D$_{5/2}$ &1.676 10$^{-2}$&1.641 10$^{-5}$\\
                               &               & 6\textit{d}$^2$7\textit{s}$^1$ & $^4$F$_{3/2}$ &1.005 10$^{-1}$&2.094 10$^{-6}$\\
6\textit{d}$^2$7\textit{s}$^1$ & $^4$F$_{7/2}$ & 6\textit{d}$^1$7\textit{s}$^2$ & $^2$D$_{3/2}$ &               &3.556 10$^{-3}$\\ 
                               &               & 6\textit{d}$^1$7\textit{s}$^2$ & $^2$D$_{5/2}$ &1.294 10$^{-1}$&1.011 10$^{-2}$\\
                               &               & 6\textit{d}$^2$7\textit{s}$^1$ & $^4$F$_{3/2}$ &               &1.858 10$^{-6}$\\
                               &               & 6\textit{d}$^2$7\textit{s}$^1$ & $^4$F$_{5/2}$ &1.294 10$^{-1}$&3.174 10$^{-5}$\\
6\textit{d}$^2$7\textit{s}$^1$ & $^4$F$_{9/2}$ & 6\textit{d}$^1$7\textit{s}$^2$ & $^2$D$_{5/2}$ &               &1.339 10$^{-2}$\\ 
                               &               & 6\textit{d}$^2$7\textit{s}$^1$ & $^4$F$_{5/2}$ &               &1.077 10$^{-4}$\\
                               &               & 6\textit{d}$^2$7\textit{s}$^1$ & $^4$F$_{7/2}$ &4.509 10$^{-1}$&1.832 10$^{-5}$\\
\hline\hline
\end{tabular}
\end{table}  

We report in \autoref{table6} the Einstein coefficients that are obtained from the electric-quadrupole (E2) and magnetic-dipole (M1) transition probabilities for the Rf$^+$ ion, using the effective Hamiltonian. The intra-configuration $^2$D$_{3/2}$ $\rightarrow$ $^2$D$_{5/2}$ transition is magnetic-dipole allowed. By using the Supplementary Material Equation S4 as function of the tabulated Einstein coefficient we obtain a lifetime of 0.7 seconds for the excited $^2$D$_{5/2}$ state (6\textit{d}$^1$7\textit{s}$^2$). The intra-configuration $^4$F $\rightarrow$ $^4$F transitions are also magnetic-dipole allowed, with the corresponding Einstein coefficients on the order of 0.1 s$^{-1}$. The inter-configuration $^2$D$_{3/2}$ $\rightarrow$ $^4F$ transitions are a priori electric-quadrupole allowed, while for the $^2$D$_{3/2}$ $\rightarrow$ $^4$F$_{3/2}$ the contributions from both M1 and E2 channels are more or less of the same magnitude. Based on this information, we derive a lifetime of about 20 seconds for the metastable $^4$F$_{3/2}$ state (6\textit{d}$^2$7\textit{s}$^1$). The calculated lifetimes of the $^4$F$_{5/2}$ and $^4$F$_{7/2}$ states (6\textit{d}$^2$7\textit{s}$^1$) equal 6 and 282 seconds, respectively.

\subsection{Optical Pumping Scheme}
\begin{figure*}[ht]
\epsfig{file=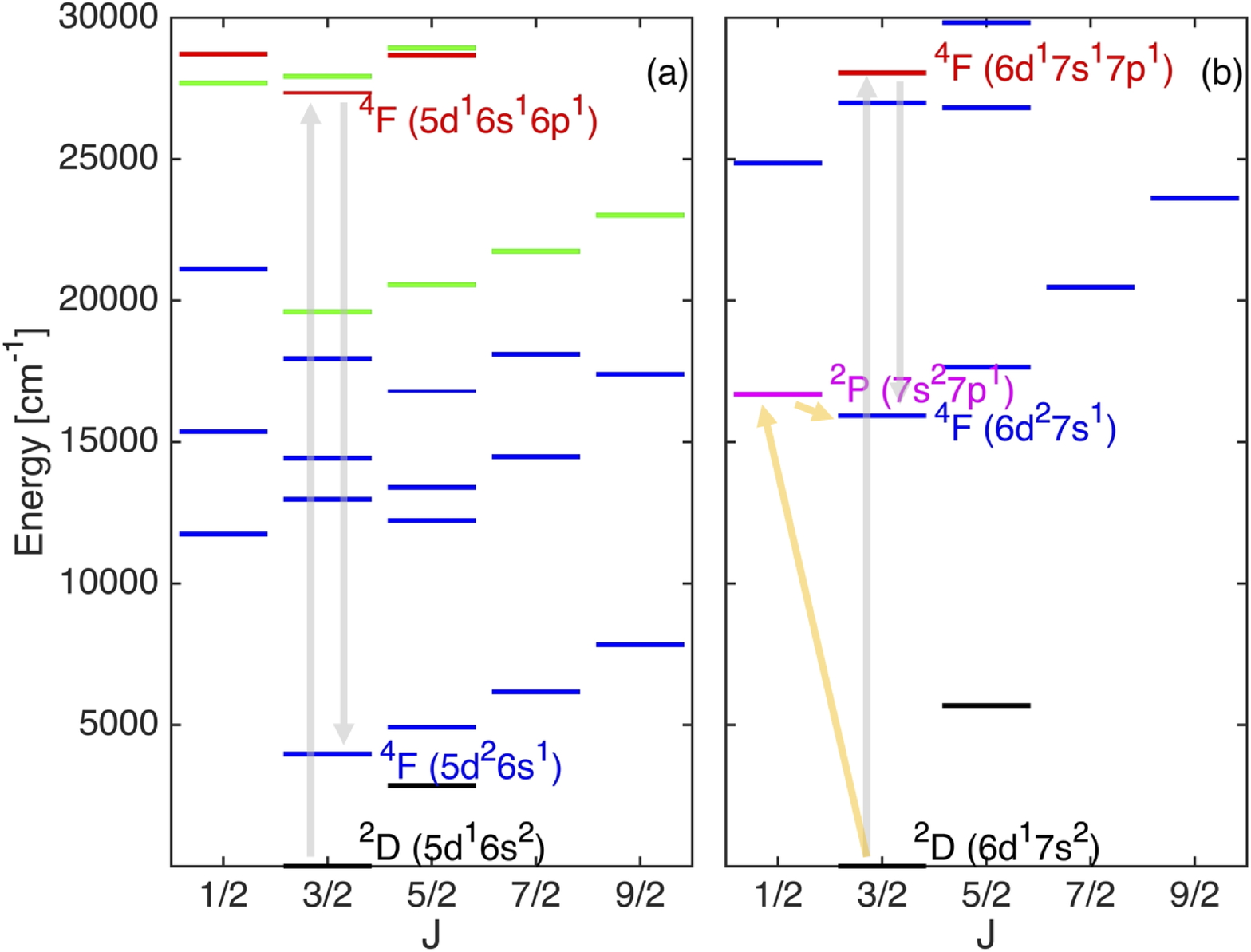, scale=0.30}
\caption{Graphical representation of selected energy levels, showing the ground and the low-lying excited states with predominant configurations (n-1)\textit{d}$^1$(n)\textit{s}$^2$ (in black); (n-1)\textit{d}$^2$(n)\textit{s}$^1$ (in blue); (n-1)\textit{d}$^3$ (in green); (n-1)\textit{d}$^1$(n)\textit{s}$^1$(n)\textit{p}$^1$ (in red); and (n)\textit{s}$^2$(n)\textit{p}$^1$ (in magenta) of Hf$^+$ (a) and Rf$^+$ (b) ions, with n = 6 and 7, respectively. Most important states are labelled for clarity and the arrows represent potential laser excitation process for the optical pumping experiment (see the text for details)}
\label{figure1}
\end{figure*}
In \autoref{figure1} we present the energy diagrams of Hf$^+$ and Rf$^+$ in the range of 0 to 30000 cm$^{-1}$. We observe a much less dense landscape of energy levels for the heavier atom, which enables the development of efficient pumping schemes for LRC experiments. For the Hf$^+$ ion, a potential LRC approach would involve pumping the ground state $^2$D$_{3/2}$ (configuration 5\textit{d}$^1$6\textit{s}$^1$) to the bright $^4$F$_{3/2}$ (5\textit{d}$^1$6\textit{s}$^1$6\textit{p}$^1$) odd parity level \cite{laatiaoui2019a,laatiaoui2020a}. The excited state radiatively decays \textit{via} two processes, reaching either the ground-state ($^4$F$_{3/2}$ (5\textit{d}$^1$6\textit{s}$^1$6\textit{p}$^1$) $\rightarrow$ $^2$D$_{3/2}$ (5\textit{d}$^1$6\textit{s}$^1$)) or the metastable state ($^4$F$_{3/2}$ (5\textit{d}$^1$6\textit{s}$^1$6\textit{p}$^1$) $\rightarrow$ $^4$F$_{3/2}$ (5\textit{d}$^2$6\textit{s}$^1$)), marked by the grey arrow in \autoref{figure1}(a) with a sizeable branching ratio (see also \autoref{table4}). Since LRC exploits ion drift in dilute gases, and because the energy separation between the metastable state and the lowest excited state of Hf$^+$ is small, we expect the metastable state to decay predominantly by collisional quenching due to competing intersystem crossing. \par
To enable LRC on the Rf$^+$ ion, we propose two different approaches based on the obtained energy levels, where the metastable state $^4$F$_{3/2}$ (configuration 6\textit{d}$^2$7\textit{s}$^1$) that has a radiative lifetime of 20 s is selectively targeted. The first approach that is marked with the grey arrow in \autoref{figure1}(b) is similar to that for Hf$^+$, where pumping the ground state $^2$D$_{3/2}$ (6\textit{d}$^1$7\textit{s}$^2$) to the bright $^4$F$_{3/2}$ (6\textit{d}$^1$7\textit{s}$^1$7\textit{p}$^1$) odd parity level effectively feeds the metastable $^4$F$_{3/2}$ (6\textit{d}$^2$7\textit{s}$^1$) state with significant branching ratio (see also \autoref{table5}). The second approach that is marked with the yellow arrow in \autoref{figure1}(b) involves pumping of the ground state to the $^2P_{1/2}$ (7\textit{s}$^2$7\textit{p}$^1$) odd parity level. In this scenario, the branching ratio to the metastable state is rather small (see \autoref{table5}). But since  our calculations predict the (7\textit{s}$^2$7\textit{p}$^1$) level to lie very close above the metastable state, we expect collisional quenching to be very efficient and to dominate the pumping process.

\section{Conclusions}
In this paper, we report a theoretical investigation of the electronic structure and spectroscopic properties of the Rf$^+$ ion. The results are obtained using the state-of-the-art 4-component relativistic multi-reference configuration interaction (MRCI) calculation. We use an effective Hamiltonian approach in conjunction with the \textit{ab initio} calculations to estimate the transition probabilities for the various inter-configurational and intra-configurational electron transitions beyond the electric-dipole approximation. We also present the energy spectrum of the lighter homologue Hf$^+$ ion. For this system, the calculated energy levels and spectroscopic properties are in good agreement with the reported experimental data, confirming the suitability of the MRCI model for this work. Thus, we expect comparable quality of the prediction for the heavy Rf$^+$ ion. \par
For Rf$^+$, the calculated energy spectrum is less dense than that obtained for the lighter homologue, suggesting electronic structure that is primarily governed by strong relativistic spin-orbit coupling interaction. Our results are consistent with the earlier studies, but we have also obtained the energy levels of the metastable states that arise from configuration 6\textit{d}$^2$7\textit{s}$^1$, and the lifetimes of the various levels. In addition, the presented method will be relevant for studying also the Rf$^+$-He interaction potential that constitutes the next step of our theoretical work. \par
Based on our calculations, we propose two possible excitation schemes to enable LRC on Rf$^+$ ion. The first scheme involves pumping the ground state $^2$D$_{3/2}$ (configuration 6\textit{d}$^2$7\textit{s}$^1$) to the bright excited state $^4$F$_{3/2}$ (configuration 6\textit{d}$^1$7\textit{s}$^1$7\textit{p}$^1$) in the ultra-violet energy range (330 nm), which effectively feeds the lowest metastable $^4$F$_{3/2}$ state (configuration 6\textit{d}$^2$7\textit{s}$^1$). The second involves pumping the ground state to the bright excited state $^2$P$_{1/2}$ (configuration 7\textit{s}$^2$7\textit{p}$^1$) in the visible energy range (600 nm), eventually reaching the metastable state via possible collisional quenching.

\section{Acknowledgements}
This project has received funding from the European Research Council (ERC) under the European Union’s Horizon 2020 research and innovation programme (Grant Agreement No. 819957). The authors gratefully acknowledge the Center for Information Technology of the University of Groningen (Peregrine), the Johannes Gutenberg University of Mainz (Mogon) and the high performance computing (HPC) group of GSI for HPC computer time.

\end{document}